\documentclass[twocolumn,pre,showpacs]{revtex4}
\usepackage{graphicx}

\begin{document}

\title{Optimizing the ensemble for equilibration in broad-histogram Monte Carlo simulations}

\author{Simon Trebst$^{1,2}$,  David A.~Huse$^{3}$, Matthias Troyer$^{1,2}$}

\affiliation{$^{(1)}$Theoretische Physik, Eidgen\"ossische Technische Hochschule
 Z\"urich, CH-8093 Z\"urich, Switzerland}
\affiliation{$^{(2)}$Computational Laboratory,  Eidgen\"ossische Technische Hochschule
 Z\"urich, CH-8092 Z\"urich, Switzerland}
\affiliation{$^{(3)}$Department of Physics, Princeton University, Princeton, NJ 08544, USA}
\date{\today}

\begin{abstract}
We present an adaptive algorithm which optimizes the
statistical-mechanical ensemble in a generalized broad-histogram
Monte Carlo simulation to maximize the system's rate of round
trips in total energy.
The scaling of the mean round-trip time from the ground state to
the maximum entropy state for this local-update method is found to
be $O([N \log N]^2)$ for both the ferromagnetic and the fully
frustrated 2D Ising model with $N$ spins.  Our new algorithm
thereby substantially outperforms flat-histogram methods such as
the Wang-Landau algorithm.
% we demonstrate a speedup by a factor of 50.
\end{abstract}


\maketitle

% Introduction ---------------------------------------------------------------------

\section{Introduction}
At first-order phase transitions and in systems with many local
minima of the free energy such as frustrated magnets or spin
glasses, conventional Monte Carlo methods simulating canonical
ensembles have very long equilibration times.
Several simulation methods have been developed to speed up
such systems, including the multicanonical method
\cite{Multicanonical}, broad histograms \cite{BroadHistograms},
parallel tempering \cite{Tempering}, the Wang-Landau algorithm
\cite{WangLandau} and variations thereof \cite{WLVariations}.
Most of these methods simulate a flat-histogram ensemble: Instead
of sampling a configuration of energy $E$ with Boltzmann weight
$w(E)\propto \exp(-\beta E)$, they use weights
$w(E) \propto 1/g(E)$, where $g(E)$ is the density of states. The
probability distribution of the energy, $n(E)=w(E)g(E)$, then
becomes constant, producing a flat energy histogram.
Naively, one might assume that sampling all energies equally often
produces an unbiased random walk in energy. However, it was
recently shown \cite{Limitations} that the growth with the number
of spins $N$ of the ``tunneling times'' between low and high
energy in any local-update flat-histogram method is stronger than 
the naive $N^2$ of an unbiased random walk in energy for various 
2D Ising models: as $\sim N^{2.4}$ for the ferromagnetic and 
$\sim N^{2.9}$ for the fully-frustrated models. For the 2D $\pm J$ 
spin glass, exponential growth was observed \cite{Limitations}.

In view of these results for flat-histogram simulations
\cite{Limitations}  we have asked how this type of simulation can
be improved, both in terms of computer time and statistical
errors. The general type of application we have in mind is to the
equilibrium behavior of a system that is very slow to equilibrate
in a conventional simulation, such as domain walls in ordered 
phases, low-energy configurations of frustrated systems 
or a spin-glass ordered phase.
Our new algorithm instead simulates a broad-histogram
ensemble, where the system can, at ``equilibrium'' in this
ensemble, wander to part of its phase diagram where equilibration
is rapid.  We look specifically at histograms that are broad in
energy, but in general another variable other than the energy
could be used.  To minimize the statistical errors of measurements
in the energy range of interest, one maximizes the number of {\it
statistically-independent} visits.  For a glassy phase, the system
will relax very little as long as it remains in that phase, so to
get a new statistically-independent visit to the phase the system
has to leave it and equilibrate elsewhere (usually at high
energies).  Thus the quantity we want our simulation to maximize
is the number of round trips -- between low and high energy --
per unit computer time.  This should minimize both the simulation's
equilibration time and the statistical errors in the low-energy
regime of interest. 

In this paper, we present an algorithm
that systematically optimizes the ensemble simulated
to maximize the rate of round trips in energy.
We use a feedback loop that reweights the ensemble based on
preceding measurements of the local diffusivity of the total
energy.  This detects the ``bottlenecks'' in the simulation as
minima in the diffusivity (at critical points in the cases we
study), and reallocates resources to those energies in order to
minimize the slowdown.
We find that the resulting statistical errors in the density of states as
estimated by this new algorithm are nearly uniform in energy,
in strong contrast to flat-histogram simulations where the errors
are much larger at low energy than at high energy.
While our algorithm is rather general and should be widely
applicable to study complex systems, we have developed and
tested it on ferromagnetic (FMI) and fully-frustrated (FFI)
square-lattice Ising models.

% Feedback for speed - the idea ----------------------------------------------

\section{Feedback of local diffusivity}
In our simulations, the system's energy does a random walk in the 
energy range between two extremal energies, $E_-\leq E \leq E_+$ 
where we take the lowest energy $E_-$ to be the system's ground 
state, although this is not necessary for our approach. 
Consider a general ensemble, with weights $w(E)$, which define
the acceptance probabilities for moves based on the Metropolis 
scheme
\begin{equation}
  p(E \rightarrow E') = \min \left( 1, \frac{w(E')}{w(E)} \right)~.
\end{equation}
Our algorithm iteratively collects data from batch runs which simulate
with a fixed ensemble. During a simulation detailed
balance is strictly satisfied at all times. For a reasonably large number 
of sweeps we can thus measure the equilibrium distribution of the 
energy in this ensemble which is $n_w(E) \propto w(E)g(E)$.
The simulated system does a biased and Markovian random walk in
configuration space. Since we bias this walk based only on total
energy, the projection of this random walk onto that variable
is what we will discuss. This projection, which ignores all
properties of the state other than its total energy, results in a
random walker that is non-Markovian, with its memory stored in the
system's configuration. Thus the simulation may be viewed as a
biased non-Markovian random walker moving along the allowed
energy range between the two extremal energies.

To measure the round trips we add a {\em label} to the walker that
says which of the two extremal energies it has visited most
recently.
The two extrema act as ``reflecting'' and absorbing boundaries for
the labeled walker: e.g., if the label is plus, a visit to $E_+$
does not change the label, so this is a ``reflecting'' boundary.
However, a visit to $E_-$ does change the label, so the plus
walker is absorbed at that boundary. The steady-state
distributions of the labeled walkers satisfy
$n_-(E)+n_+(E)=n_w(E)$.
It is important to note that the behavior of the labeled walker is 
{\em not} affected by its label except when it visits one of the extrema 
and the label changes. 
When the unlabeled walker is at equilibrium, the labeled
walker is in a nonequilibrium steady state. Let $f(E) =
n_+(E)/n_w(E)$ be the fraction of the walkers at $E$ that have
label plus, so they have most recently visited $E_+$. The
above-discussed boundary conditions dictate $f(E_-)=0$ and
$f(E_+)=1$.

To calculate the rate of round trips, we note that in steady state
the current $j$ of the labeled walkers is independent of $E$. The
plus and minus walkers drift in opposite directions and
the equilibrium {\it unlabeled} walker has no net current.
We first examine the case of a continuous energy $E$. 
The steady-state current from $E_+$ to $E_-$ to first order in the 
derivative is
\begin{equation}
j=D(E)n_w(E){{df}\over{dE}}~, \label{Eq:Current}
\end{equation}
where $D(E)$ is the walker's diffusivity at energy $E$.  There is
no current if $f$ is constant, since this is equilibrium; this is
why the current is to leading order proportional to $df/dE$.
If one rearranges the above equation and integrates on both sides,
noting that $j$ is a constant and $f$ runs from 0 to 1, one
obtains
\begin{equation}
{1\over{j}}=\int_{E_-}^{E_+}{{dE}\over{D(E)n_w(E)}}~.
\label{Eq:1overj}
\end{equation}
In the following we separately discuss how we can maximize the 
rate of round-trips for Metropolis and $N$-fold way dynamics based
on this estimate of the current.

% Metropolis
\subsection{Metropolis dynamics}

For Metropolis dynamics the rate of round-trips is simply proportional
to the current. To maximize the round-trip rate, the above integral,
Eq. (\ref{Eq:1overj}), must be minimized. However, there is a constraint:
$n_w(E)$ is a probability distribution and must remain normalized.
We do this by adding a Lagrange multiplier:
\begin{equation}
\int_{E_-}^{E_+}dE\left({1\over{D(E)n_w(E)}}+\lambda n_w(E)\right)~.
\label{Eq:Integrand}
\end{equation}
To minimize this integrand, the ensemble, that is the weights
$w(E)$ and thus $n_w(E)$ are varied. At this point we assume that
the dependence of $D(E)$ on the weights can be neglected.
This is justified by noting that the rates of transitions between
configurations depend only on the {\it ratios} of weights, so the diffusivity
$D(E)$ is unchanged when the weights are multiplied by an
energy-independent constant.  By ignoring the variation of $D(E)$
with the weights, we are assuming that the adjustments to the
weights are slowly-varying in energy, which is true for most
systems, particularly for large systems where the energy range
being studied is large. With this assumption, the optimal
weighting which minimizes the above integrand is
\begin{equation}
n_w^{(opt)}={1\over{\sqrt{D(E)\lambda}}}={{df^{(opt)}}\over{dE}}~.
\end{equation}
Thus for the optimal ensemble with Metropolis dynamics, the probability 
distribution is simply inversely proportional to the square root of the local
diffusivity.

% N-fold way, "two clocks"
\subsection{$N$-fold way dynamics}
Since Metropolis dynamics can be slowed down by high rejection rates
of singular moves, e.g. in the vicinity of the fully polarized ground state of 
the FMI, or the occurrence of multiple, generally accepted zero-energy 
moves  it can be advantageous to introduce rejection-free single-spin flip
updates such as the $N$-fold way  \cite{NFoldWay}.
$N$-fold way dynamics involve two time scales, the walker's time
and the computer time. At a given energy level the two time scales 
differ by the (energy-dependent) lifetime of a given spin configuration. 
%With the ratio $t(E)$ we denote the amount of computer time spent at 
%an energy level $E$ per unit walker's time spent at $E$. 
The random walk with $N$-fold way dynamics is an equilibrium process 
when measured in walker's time, that is the equilibrium distribution, 
$n_w(E)$, is proportional to the amount of {\em walker's} time the walker
spends at $E$. 
However, for the ensemble optimization with $N$-fold way dynamics we
want to speedup the random walk measured in {\em computer} time. 
This setup with two clocks requires a  slightly different reweighting procedure
than is presented above for Metropolis dynamics.

As for the Metropolis dynamics the amount of walker's time it takes to make 
a round trip is proportional to $1/j$ given in Eq. (\ref{Eq:1overj}).  However, 
we are interested in minimizing the amount of computer time spent, so we
need to multiply this by the ratio of computer time to walker's time at $E$ which
we denote as $t(E)$.  
Let us assume the distribution $n_w(E)$ is normalized to integrate to one.  
Then for one unit of walker's time, the fraction spent at $E$ in $dE$ is $n_w(E)dE$.  
The amount of computer time used per unit walker's time is thus
\begin{equation}
  T=\int_{E_-}^{E_+}n_w(E)t(E)dE~.
\end{equation}
To find the weights that minimize the round-trip time as measured in
computer time, the full quantity we want to minimize is thus
\begin{equation}
  \int_{E_-}^{E_+}{{dE}\over{D(E)n_w(E)}}\int_{E_-}^{E_+}t(E')n_w(E')dE'~.
  \label{Eq:integrand}
\end{equation}
Since the probability distribution $n_w(E)$ occurs in both the numerator and
denominator of the integrand there is no need to enforce its normalization by a
Lagrange multiplier. To extremize the integrand, we will again vary the 
weights $w(E)$, which gives the following condition for the optimum:
\begin{equation}
  {{T}\over{D(E)n_w^2(E)}}={{t(E)}\over{j}}~,
\end{equation}
so the weights should be chosen to give the optimal distribution
\begin{equation}
  n_w^{(opt)}(E)=\sqrt{{{j T}\over{D(E)t(E) }}}~.
\end{equation}
For the optimal ensemble with $N$-fold way dynamics, the probability
distribution is thus larger at the points with smaller $t(E)$ (since they do 
not cost a lot of computer time) and smaller diffusivity $D(E)$.

\subsection{Feedback iteration}
To feed back we simulate with some trial weights $w(E)$, get
steady-state data for $n_w(E)$ and $f(E)$ and thus obtain estimates
for the diffusivity via
\begin{equation}
D(E)={{j}\over{n_w(E){{df}\over{dE}}}}~.
\end{equation}
For Metropolis dynamics chose new weights $w'(E)$ so that
\begin{equation}
n_{w'}(E)=A\sqrt{n_w(E){{df}\over{dE}} }~,
\end{equation}
where $A$ is a normalization constant whose value is not needed to
run the next ``batch'' of the simulation with the new weights $w'(E)$. 
For $N$-fold way dynamics the new weights $w'(E)$ are chosen to be
\begin{equation}
n_{w'}(E)=\sqrt{n_w(E){{df}\over{dE}} {T\over{t(E)}}}~.
\end{equation}
In practice we work with the logarithms of the weights, so the reweighting 
becomes
\begin{eqnarray}
  \ln w'(E) & = & \ln w(E) \\ \nonumber
  & & + \frac{1}{2} \left( \ln\left\{{{df}\over{dE}}\right\} - \ln n_w(E) - \ln t(E) \right) \;,
\end{eqnarray}
where all energy independent terms have been dropped as they introduce a constant
shift only. For Metropolis updates the last term $\ln t(E)$ can also be dropped.
Each subsequent batch should be run significantly longer than the
previous one -- in our implementation we double the number of sweeps -- 
in order to get better statistics, and fed back to improve the estimates 
of the optimal weights.

%--- Numerical implementation -------------------------------------

\section{Implementation and Applications}
We implemented this algorithm for 2D Ising models with
single-spin-flip Metropolis and $N$-fold way dynamics,
found the optimal ensembles for the FMI and FFI models,
obtained the scaling of round-trip times and
calculated the density of states and its statistical
errors for both models.  In both cases we used the ground state,
$E_-=E_0$, and zero energy, $E_+=0$, as the energy limits.

% FFI: histograms / diffusivity
\begin{figure}[t]
  \includegraphics[scale=0.35]{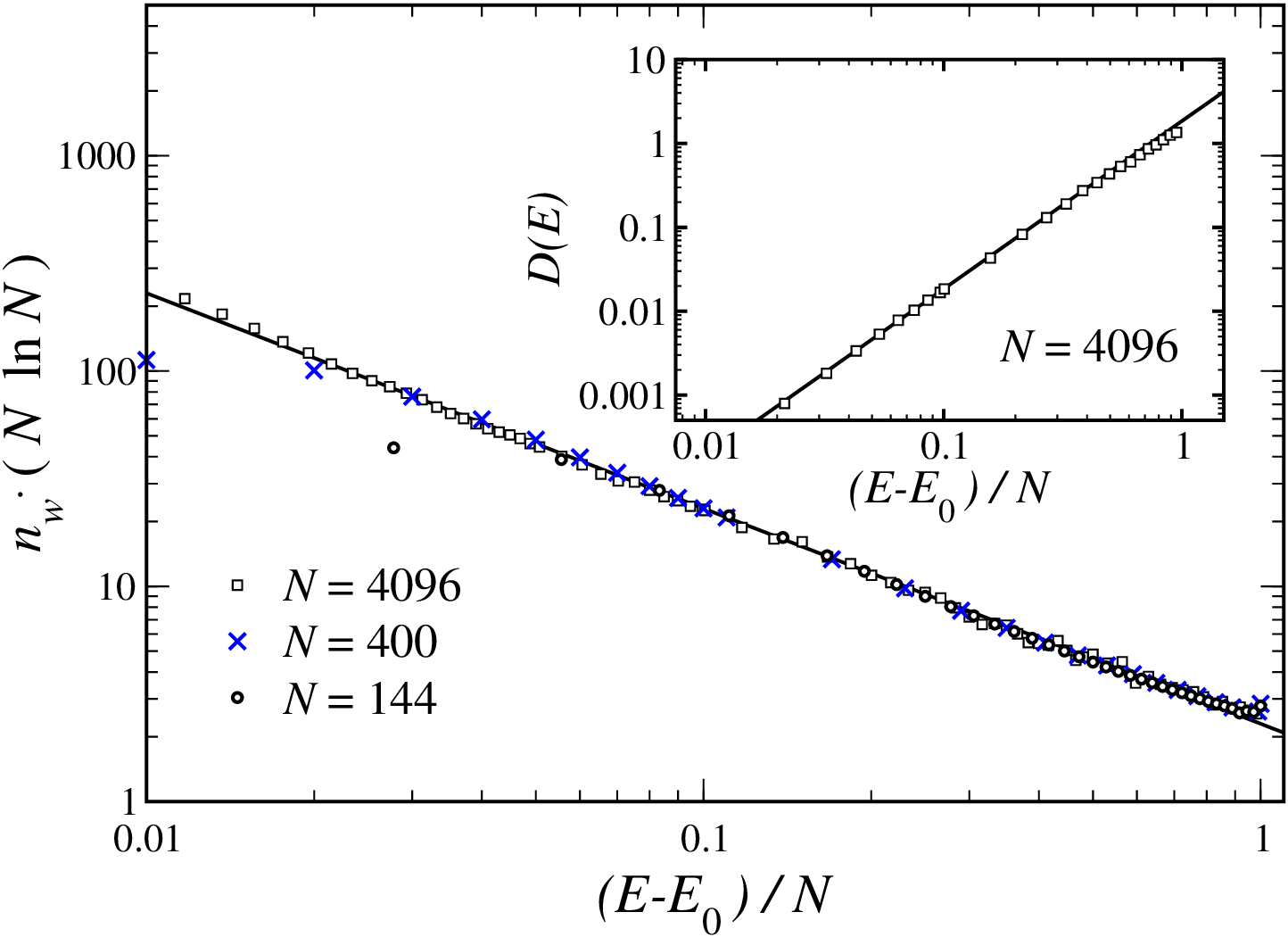}
  \caption{Histograms of the optimal ensemble for the 2D fully
        frustrated Ising model with Metropolis dynamics.
              For various system sizes and a broad energy range the
              rescaled data points collapse onto a a power-law
              divergence $[(E-E_0)/N\ln N]^{-1}$ (bold line).
                   The inset shows the diffusivity $D(E)$ for the same model
                   which is proportional to $[(E-E_0)/N]^2$ (bold line).
                   }
  \label{Fig:FFI_Histograms}
\end{figure}

% fraction of labeled walkers
In the initial batch mode step we simulated a flat-histogram
ensemble for small system sizes using either the exact density 
of states \cite{footnote} or a rough estimate thereof calculated with the 
Wang-Landau algorithm \cite{WangLandau}. For larger systems 
($N>64\times64$) where the round-trip times for the flat-histogram 
ensemble are more than a magnitude larger than for the optimized 
ensemble, we produced an initial estimate of the optimal weights by 
extrapolating the optimized weights of smaller systems.
For all batch mode steps the fraction of labeled walkers, $f(E)$, was
determined by recording two histograms, one for the equilibrium
(unlabeled) walker and one for the labeled walker which most
recently visited $E_-$. The derivative $df/dE$ was then estimated by a 
linear regression of several neighboring points at each energy level. 
The number of points used for the regression can be reduced for 
subsequent batch mode steps as the estimate of $f(E)$ becomes increasingly
accurate due to better statistics. In the final batch mode steps of our 
calculations the regression was performed using a minimum of three points.
In general, there is a trade-off between the accuracy in the measurement of 
the local diffusivity and the number of feedback steps. 
For the Ising models we study we found rapid convergence to the optimal
ensemble. For small systems, $N\leq32\times32$, an initial batch mode step
with some $10^5$ to $10^6$ sweeps was sufficient to find the optimal weights
after the first feedback step.
Since the possible energy levels are discrete for the Ising model, special 
care is taken when applying the reweighting derived for the continuum limit, 
particularly at the boundaries of the energy interval $[E_-,E_+]$.
However, we did not encounter any subtlety for either model.

\subsection{Fully frustated Ising model}

% FFI: scaling of round-trip times
\begin{figure}[t]
  \includegraphics[scale=0.35]{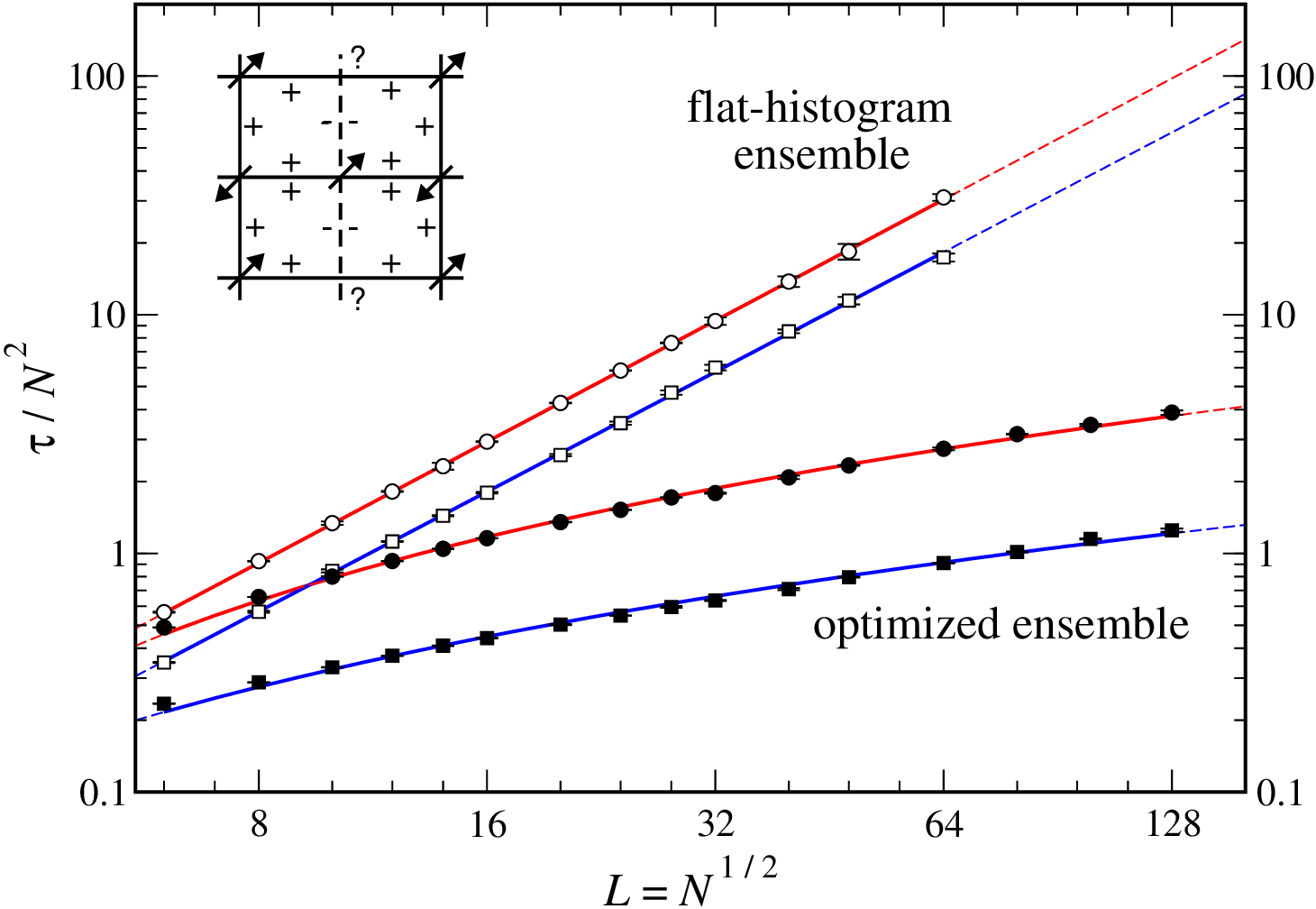}
  \caption{Scaling of round-trip times in the energy interval
        $[-N, 0]$ for the flat-histogram (open symbols) and optimized
        ensemble  (filled symbols) of the 2D fully frustrated Ising model
        with Metropolis (circles) and $N$-fold way (squares) dynamics.
        The solid lines correspond to a logarithmic (power-law) fit
        for the optimized (flat-histogram) ensemble.
        The inset illustrates the frustrated couplings of the fully
        frustrated model.}
  \label{Fig:FFI_Scaling}
\end{figure}

% histograms of the optimized ensemble
We first present our results for the fully-frustated model (FFI),
which has a critical point at its ground state, and shows rather
simple scaling of our algorithm's behavior with energy and system
size. For the optimized ensemble of the FFI the histogram of the
equilibrium random walker is no longer flat, but exhibits
a power-law divergence at its ground state, as shown in
Fig.~\ref{Fig:FFI_Histograms}.
This divergence reflects the behavior, $D(E)\sim [(E-E_0)/N]^2$
of the diffusivity, as is seen in the inset of Fig.~\ref{Fig:FFI_Histograms}.
These power-law behaviors extend from the first few points,
$E-E_0=O(1)$, up nearly to zero energy, $E-E_0=O(N)$.
If we accept that the critical exponent for the diffusivity is indeed 2,
then the optimal distribution scales as $n_w\sim 1/[(E-E_0)\ln N]$,
and the round-trip time as $\tau\sim(N\ln N)^2$, consistent with
our results shown in Figs.~\ref{Fig:FFI_Histograms} and
\ref{Fig:FFI_Scaling}.

% N-fold way
Noting that for our optimized ensemble the system spends a large
fraction of its time near the ground state where many Metropolis
moves are rejected, we applied a version of our algorithm
that instead uses single-spin-flip rejection-free $N$-fold way
updates. We find the $N$-fold way updates do give a significant
speedup compared to Metropolis dynamics, but do not change the 
$\tau\sim (N\ln N)^2$ scaling of the round-trip time.
In comparison to the performance of flat-histogram sampling 
we find a substantial speedup up to a factor of around $50$ for 
the largest simulated system with $N=128\times128$ spins,
see Fig.~\ref{Fig:FFI_Scaling}.

%--- FMI -------------------------------------

\subsection{Ferromagnetic Ising model}

% FMI: histograms / fraction
\begin{figure}[t]
  \includegraphics[scale=0.35]{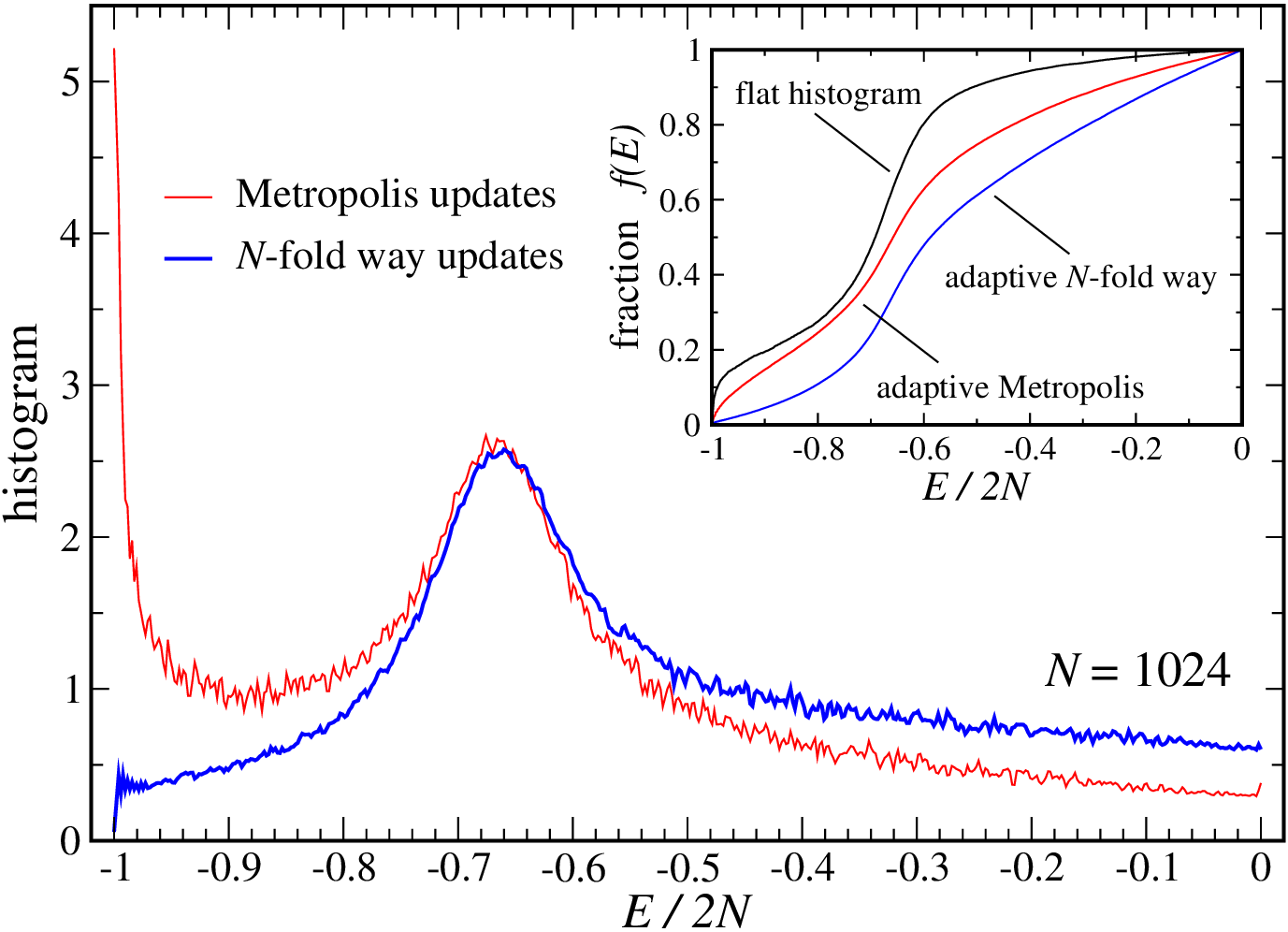}
  \caption{Histograms for the 2D ferromagnetic Ising model obtained after feedback:
        		for the optimal ensemble a peak evolves around the critical energy in the histograms.
       		The additional peak near the fully polarized groundstate found for Metropolis updates
                   (thin line) can be eliminated by changing the dynamics to $N$-fold way updates
                	(bold line).
                  The inset shows the fraction $f(E)$ of walkers which have most recently visited
                  $E_+=0$ for flat-histogram (multicanonical) sampling and the optimal ensembles
                  for Metropolis and $N$-fold way dynamics.}
  \label{Fig:FMI_Histograms}
\end{figure}

\begin{figure}[t]
  \includegraphics[scale=0.35]{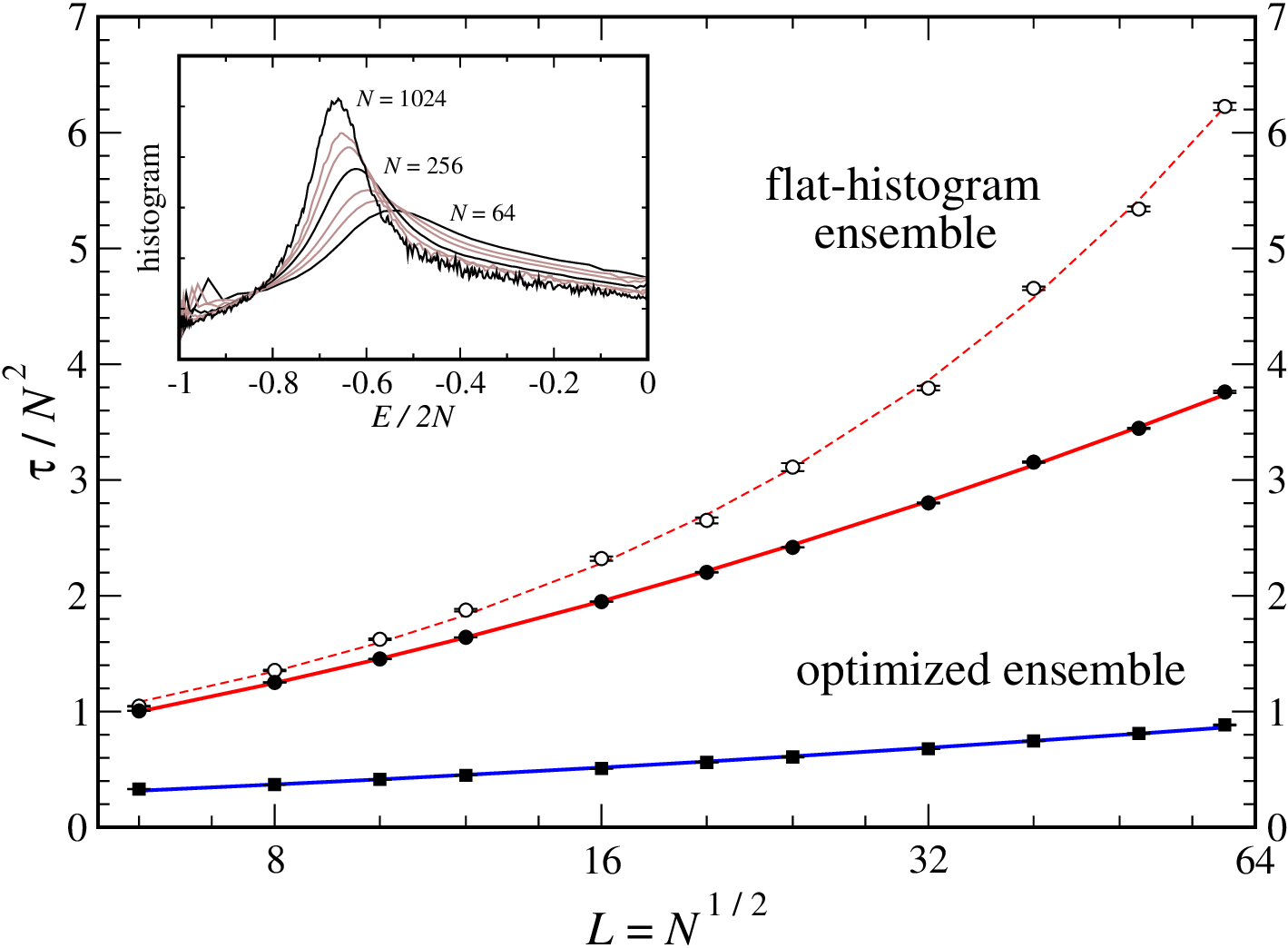}
  \caption{Scaling of round-trip times in the energy interval
        $[-2N, 0]$ for the flat-histogram ensemble (open symbols)
        and the optimal ensemble (filled symbols) of the 2D
        ferromagnetic Ising model with Metropolis (circles)
        and $N$-fold way dynamics (squares).
        The solid (dashed) lines correspond to a logarithmic 
        (power-law) fit for the optimized (flat-histogram) ensemble.
        The inset shows the scaling of histograms for the optimal
        ensemble for $N$-fold way updates.
         }
  \label{Fig:FMI_Scaling}
\end{figure}

We now turn to the results for the ferromagnetic Ising model (FMI)
which exhibits a finite-temperature second order phase transition.
After applying the feedback, we find a peak in the histogram near
the critical energy, as shown in Fig.~\ref{Fig:FMI_Histograms}.
For Metropolis updates a second divergence close to the fully
polarized ground state appears which is eliminated by changing the
dynamics to rejection-free $N$-fold way moves. However, the
minimum in the diffusivity at the critical point remains with
$N$-fold way dynamics and the resulting peak in the histogram is
not suppressed. With increasing system size this power-law
divergence moves towards the critical energy of the infinite
system, $E_c/2N \cong -0.71$, as illustrated in the inset of
Fig.~\ref{Fig:FMI_Scaling}. For both types of single-spin-flip
moves we find that the rate of round trips between the
magnetically ordered and disordered phases of the ferromagnet
appear to scale as $\tau\sim(N\ln N)^2$ as for the FFI model, see
Fig.~\ref{Fig:FMI_Scaling}.

% FMI: scaling of round-trip times / histograms

% density of states / statistical errors ----------------------------------------

\section{Statistical errors}

% FMI / FFI: statistical errors
\begin{figure}[t]
  \includegraphics[scale=0.35]{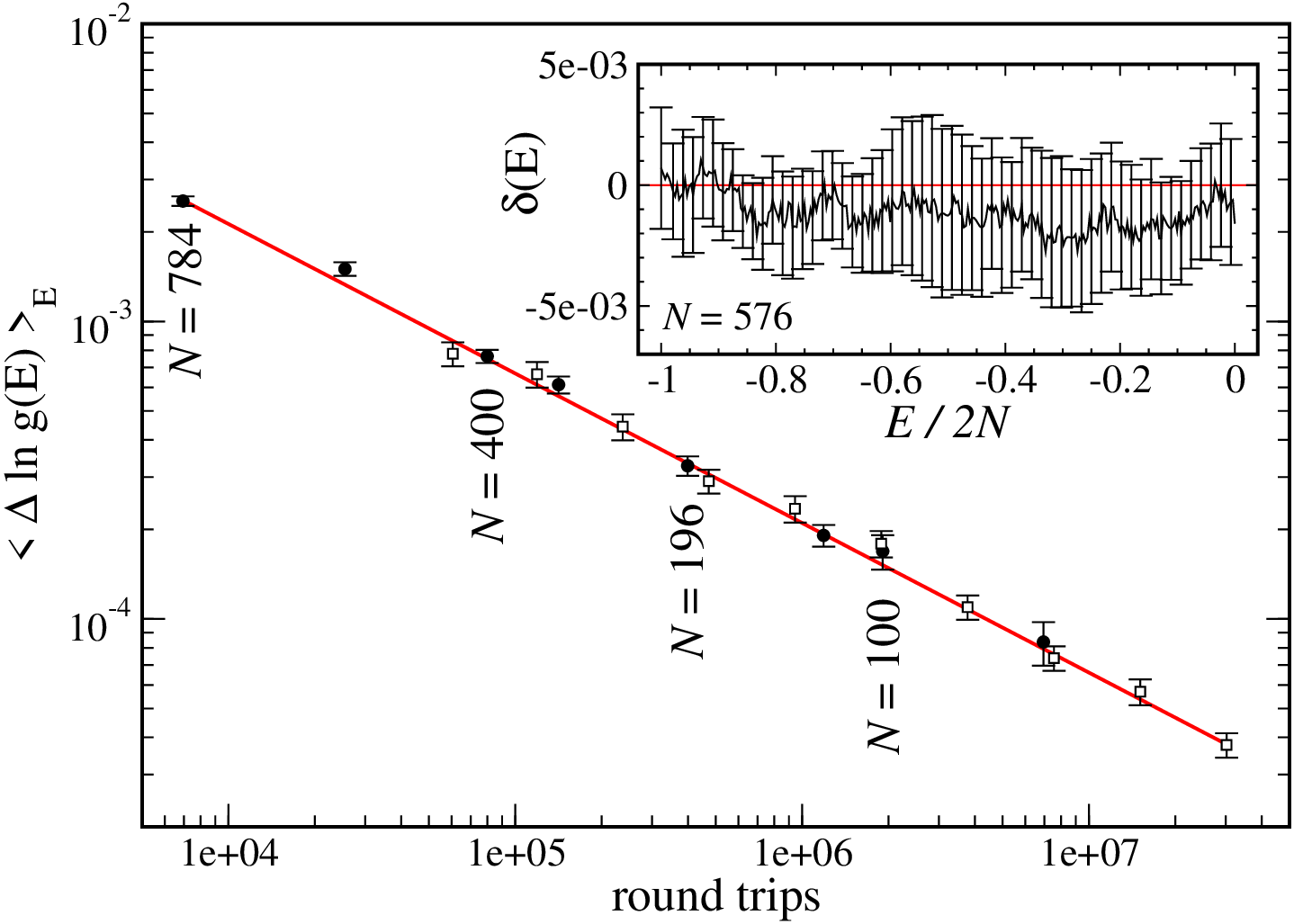}
  \caption{Average statistical error $\langle \Delta \ln g(E)\rangle_{E}$ 
  	of the computed density of states of the FFI model versus the number 
	of round trips in the energy interval $[-N, 0]$.
       	The statistical errors were obtained for 16 independent
        	runs and averaged over all energies for $N=36$ (open symbols).
       	Data points for larger system sizes are superimposed (solid symbols),
        	with the system size increasing from right to left.
        	The statistical error reduces like $1/\sqrt{\mbox{round trips}}$
        	(solid line) for all system sizes.
        	The inset shows the deviation,
        	$\delta(E) = \ln g(E) - \ln g_{\mbox{exact}}(E)$,
        	from the exact result for the $24\times24$ FMI model obtained
        	for 16 independent runs with $3.2\cdot10^6$ sweeps.}
  \label{Fig:StatisticalError}
\end{figure}

Finally we address the statistical errors of measurements performed
during the simulation. Standard tools can be used for the error analysis
as the simulated random walk in configuration space is a conventional
Markov chain Monte Carlo simulation. Only the projection of this random
walk onto energy space becomes non-Markovian which is irrelevant for
the measurements.

For each batch mode step simulating a fixed statistical ensemble $w(E)$ 
we can measure the density of states, $\ln g(E) = \ln n_w(E) - \ln w(E)$,
from the recorded equilibrium distribution $n_w(E)$. 
Comparing our results with the exact density of states we find perfect
agreement within the statistical errors as illustrated for the FMI in the inset
of Fig.~\ref{Fig:StatisticalError}. 
The observed distribution of statistical errors is nearly flat in
energy, which is a further improvement compared to flat-histogram
simulations where the errors can be orders of magnitude larger at
low energy than at high energy \cite{WangLandau}. The statistical
error is found to scale as $\Delta \ln g(E) \sim
1/\sqrt{\mbox{round trips}}$ with the number of  round trips in
energy which is shown in the main panel of
Fig.~\ref{Fig:StatisticalError}. For different system sizes we
find the statistical errors to collapse onto a single
$1/\sqrt{\mbox{round trips}}$ dependence which {\it a posteriori}
validates our goal of maximizing the rate of round trips.

% Summary / Conclusions ----------------------------------------------------

\section{Conclusions}
The presented algorithm should be widely applicable to study the 
equilibrium behavior of complex systems, such as glasses, 
dense fluids or polymers.
To speed up the system's equilibration the rate of round trips in
energy is maximized by systematically optimizing the statistical
ensemble based on measurements of the local diffusivity.
We find that the relative statistical error in the density of states as
calculated with the new method scales as
$O(1/\sqrt{\mbox{round trips}})$.
For the 2D ferromagnetic and fully frustrated Ising models the
round-trip time from the ground state to the maximum entropy state
scales like $O([N \log N]^2)$ which is a significant speedup
compared to the power law behavior $O(N^{2+z})$ of 
flat-histogram algorithms.

The idea of performing round-trips in energy is similar to the parallel 
tempering algorithm \cite{Tempering} which simulates replicas of the 
system at various temperatures. The swapping of replicas at neighboring 
temperatures can be viewed as a random walk of the replicas along the 
temperature. In order to maximize the round-trips in temperature 
one can use our algorithm to systematically optimize the simulated 
temperature set which we will discuss in a forthcoming publication 
\cite{FutureTempering}.

% Acknowledgments ---------------------------------------------------------

\section{Acknowledgments}
We thank S.~Sabhapandit for providing the exact density of states
for the FFI and R.~H.~Swendsen for helpful discussions. 
ST acknowledges support by the Swiss National Science Foundation. 
DAH is supported by the NSF through MRSEC grant DMR-0213706.

% References
\bibliography{paper}

\begin{thebibliography}{28}
\bibitem{footnote}
\bibnamefont{The exact density of states was calculated with the algorithms \cite{Beale:96}  
and \cite{Saul:94} for the FMI (up to $N=60\times60$) and FFI (up to $N=28\times28$) respectively.}
\end{thebibliography}

\end{document}